\documentclass[prc,twocolumn,showpacs]{revtex4-1}
\usepackage{graphicx}
\usepackage{amsmath, amsthm, amssymb}
\usepackage{multirow}
\usepackage{subfigure}
\usepackage{float}
\begin{document}

\title{A Scalable, Self-Analyzing Digital Locking System for use on Quantum Optics Experiments}
\author{B. M. Sparkes, H. M. Chrzanowski, D. P. Parrain, B. C. Buchler, P. K. Lam, and T. Symul}
\affiliation{ Centre for Quantum Computation and Communication Technology, Department of Quantum Science, Research School of Physics and Engineering, The Australian National University, Canberra, ACT 0200, Australia}
\begin{abstract}
Digital control of optics experiments has many advantages over analog control systems, specifically in terms of scalability, cost, flexibility, and the integration of system information into one location. We present a digital control system, freely available for download online, specifically designed for quantum optics experiments that allows for automatic and sequential re-locking of optical components. We show how the inbuilt locking analysis tools, including a white-noise network analyzer, can be used to help optimize individual locks, and verify the long term stability of the digital system. Finally, we present an example of the benefits of digital locking for quantum optics by applying the code to a specific experiment used to characterize optical Schr\"odinger cat states.
\end{abstract}

\maketitle
\section{Introduction}
\label{sec:dl_introduction}
Examples of the use of digital control algorithms, rather than analog electronics, to lock the frequency of single \cite{dl9} and multiple \cite{dl14} lasers date back to 1998. This progression has been motivated by the fact that a digital system can offer distinct advantages over its analog counterpart. These include cost savings, as features can be programmed rather than manufactured or purchased, and space savings, as most of the features can be included in the code they take up only as much space as the hardware needed to run it. Mostly, however, the advantage of digital locking over analog systems lies in its flexibility: the ability to change the functionality of the system by altering the control code rather than having to purchase new equipment or physically modify old equipment. This makes a digital system easily reproducible, as well as having the ability to add complex logic which would be difficult to include with analog circuits alone (for examples see Ref.s \cite{dl10} and \cite{dl15}). Using the added power of digital control has lead many to develop novel methods for frequency stabilizing lasers (for example see Ref.s \cite{dl13,dl12,dl11,dl19,dl17,dl20}).\\
Quantum optics is one branch of science which can benefit from digitization and computer control. Here we present a code designed for quantum optics experiments using field programmable gate arrays (FPGAs) programmed with National Instruments LabVIEW$\textregistered$ software. This code is freely available to be downloaded \cite{code} and to be modified, as both a pedagogical tool, and to allow individual users to tailor it to their specific needs.\\
\subsection{Digital Control for Quantum Optics Experiments}
\label{sec:dl_quantum_optics_experiments}
One component common to many quantum optics experiments is the cavity - for instance mode-cleaner style cavities to select only one spatial mode of an input field or Fabry-Perot cavities to select a specific frequency. Cavities are not just used for filtering, however, and can also be an integral part of an experiment. For instance, cavities containing non-linear optical crystals (known as optical parametric oscillators - OPOs) are used to produce entanglement \cite{en13} and squeezing \cite{sqz4,sqz5}, two powerful quantum resources. These effects are extremely sensitive to noise or loss in the system. Therefore the properties of the system used to lock an OPO must be stringently monitored. As well as the use of cavities, another issue of concern is the relative phase between various fields. For instance, homodyne detection using a local oscillator \cite{hom1,hom2}.\\
Quantum optics experiments can require many locking loops and sophisticated data acquisition, and therefore these experimental systems can become highly complex. As an example, in 2003 Bowen \textit{et al.} demonstrated the teleportation of the quadrature amplitudes of two light fields \cite{bowen}. To achieve this the experiment required a frequency doubling cavity, a high finesse ring cavity used to seed a pair of OPOs, as well as two homodyne detector set-ups requiring phase control.\\
More recently, another example of the complexity of quantum optics experiments is the work carried out by Yukawa \textit{et al.} in 2008 to generate four-mode cluster states for use in quantum computing \cite{furusawa}. This experiment required a frequency doubler, which was used to pump four OPOs, and four homodyne set-ups were required for the measurements. From these examples it can be seen that a digital control system for quantum optics experiment would need to be extendible to many locks, flexible in the style of locks it can control, as well as be as effective, or better, than current analog controller in reducing noise.\\
As more complex experiments are developed to investigate further into the quantum realm, the practicality of digital control will become more apparent. This is because digital control allows for integration of all relevant information about the system to be accessed, and controlled, from one location. This, in turn, allows the system to take into account the sequential nature of the locks comprizing the experiment. Also, the complete system information can be used for conditional acquisition, to ensure that data is only recorded when the system is in the desired state.\\
\subsection{Control Theory}
\label{sec:dl_control_theory}
\begin{figure}
\begin{center}
\includegraphics[width=\columnwidth]
{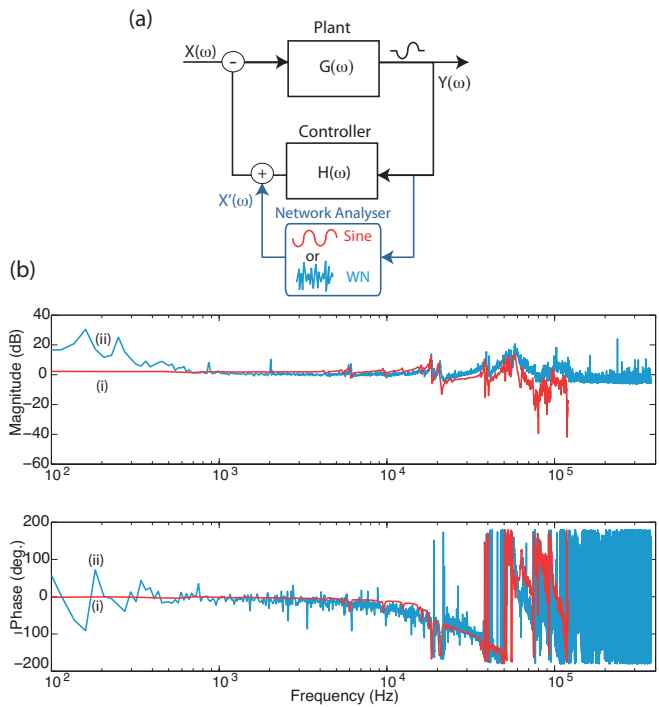} 
\caption{(a) Block diagram of a basic feedback control system. Here $G(\omega)$ and $H(\omega)$ are the transfer functions of the plant and controller respectively, with $X(\omega)$ the input noise and $Y(\omega)$ the output of the system (i.e. error signal). Also shown is the placement of a network analyzer, with $X'(\omega)$ the extra noise added to the system in the form of a swept sine wave (Sine) or white noise (WN). (b) Bode plot of the transfer function of a code-cleaner cavity for (i) a network analyzer (model MS4630B from Anritsu) and (ii) a white noise generator included in the code.}
\label{fig:dl_controltheory}
\end{center}
\end{figure}
This section provides an overview of the relevant control theory, for a more in depth description see, for example, Ref. \cite{dorf}. Fig. \ref{fig:dl_controltheory}(a) shows the basics of a closed loop feedback system. A plant - the system to be controlled - having a transfer function (i.e. frequency response) $G(\omega)$ produces some form of error signal $Y(\omega)$. This is passed to the controller, having a transfer function $H(\omega)$, used to suppress the noise added to the system $X(\omega)$. The system transfer function is defined as
\begin{eqnarray}
T(\omega) & = & \frac{Y(\omega)}{X(\omega)} \nonumber \\
& = & \frac{G(\omega)}{1 + GH(\omega)},
\label{eq:dl_transferfunction}
\end{eqnarray}
where we take the shorthand $GH(\omega) \equiv G(\omega)H(\omega)$. For quantum optics experiments, as with all locking servo applications, we are interested in three properties of the controller: how well it can suppress noise at low frequencies; what frequency range it can achieve this over; and how stable the lock is. Here we define the bandwidth of the system to be the frequency $\omega_B$ at which the controller no longer suppresses noise, i.e. the unity gain frequency $GH(\omega_B)=1$. The stability of the system is usually measured using the Nyquist stability criterion \cite{fc7} using the phase margin (PM) defined by
\begin{equation}
PM = \pi + \angle GH(\omega_B).
\label{eq:dl_phasemargin}
\end{equation}
That is, the clearance of the phase of $GH(\omega_B)$ from $-\pi$.\\
One way of investigating the system transfer function is to use a network analyzer to add known noise $X'(\omega) \gg X(\omega)$ to the system and measuring the frequency response. For commercial network analysers this extra noise is usually in the form of a swept sine wave, though white noise - that is noise with the same amplitude at all frequencies - can also be used. The placement of a network analyzer to measure the system transfer function is shown in Fig. \ref{fig:dl_controltheory}(a), and examples of magnitude and phase (i.e. Bode) plots for an unlocked mode-cleaner cavity, measured using this method, are shown in Fig. \ref{fig:dl_controltheory}(b).\\
Another measure of the efficacy of a lock is  the deviation of the error signal from its desired value over time. This is usually measured using the root mean squared (RMS) method, which is expressed mathematically as
\begin{equation}
RMS = \sqrt{\frac{\sum_{i=1}^{n} \left(S_i - S_{des}\right)^2}{n}}.
\label{eq:dl_rms}
\end{equation}
Here $S_i$ is the value of the error signal at a particular time $i$, $S_{des}$ is the desired value of the error signal (usually 0), and $n$ is the number of points the RMS was measured over.
\section{Digital Controller Design}
\label{sec:dl_code}
The digital control system presented here is based on the Pound-Drever-Hall locking technique \cite{l6,l7} that uses modulation of the laser frequency to produce an anti-symmetric error signal (see Fig. \ref{fig:apdl_setup}(a) inset) which is then fed back to the actuating mechanism to keep the system locked to a desired position. The code allows for automatic, sequential locking, as well as analysis of the locks comprising the experiment to facilitate optimizing procedures.
\subsection{Hardware}
\label{sec:dl_hardware}
\begin{figure}
\begin{center}
\includegraphics[width=\columnwidth]
{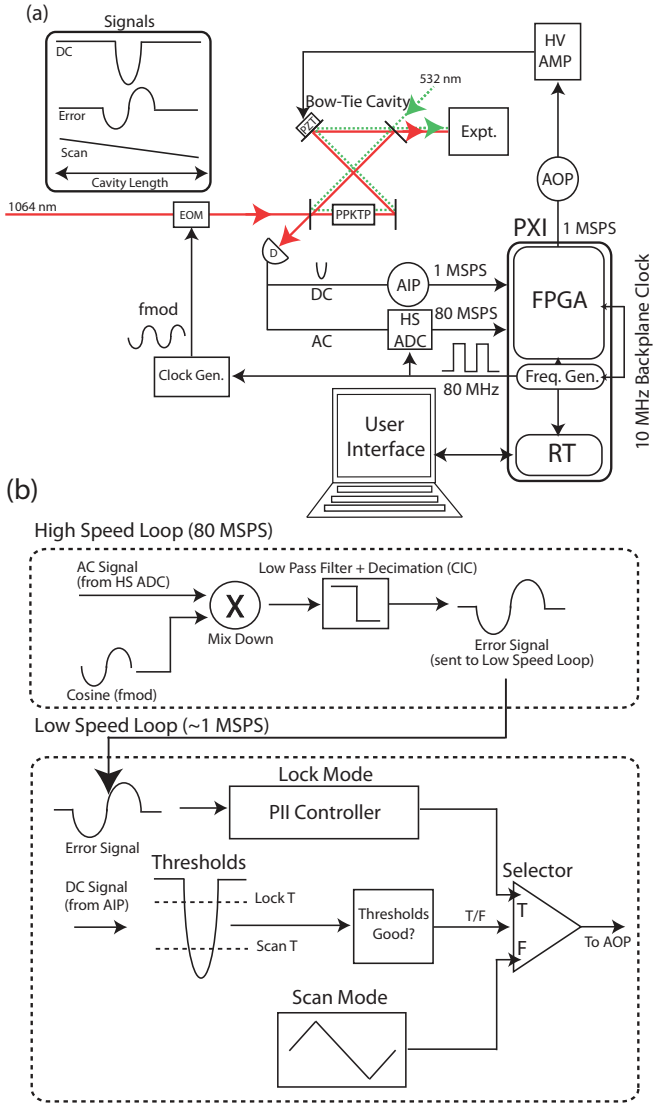} 
\caption{(a) Depiction of the physical system and hardware required for digital locking using the code presented here, in this case for a bow-tie OPO cavity doubly resonant with a green (532 nm) and a red (1064 nm) field, the latter being used to lock the cavity position on reflection. MSPS - M Samples s$^{-1}$; PXI - PXI chassis; FPGA - field programmable gate array card; Freq. Gen. - frequency generator; RT - real time controller; AIP - analog input to FPGA (sampling frequency $<$ 1 MSPS); AOP - analog output from FPGA (sampling frequency $<$ 1 MSPS); HS ADC - high speed analog to digital converter (80 MSPS); D - detector; Clock Gen. - digital clock generation board; fmod - modulation frequency for lock; EOM - electro-optic modulator; PZT - piezo-electric transducer; PPKTP - periodically poled potassium titanyl phosphate crystal; HV Amp - high voltage amplifier; and Expt. - rest of experiment. Inset shows the different signals which are used for locking (DC - dc reflection signal; Error - Error signal; Scan - scan function) as a function of PZT position. (b) Depiction of LabVIEW$\textregistered$ code used to program FPGA cards. CIC - Cascaded-Integrator-Comb filter; PII controller - proportional, integral, double integral controller; Scan T - scan threshold; Lock T - lock threshold; T - true; F- false.}
\label{fig:apdl_setup}
\end{center}
\end{figure}
Fig. \ref{fig:apdl_setup}(a) shows the basic setup for a single lock, in this case for an OPO. A frequency generation (FG) signal at 80 MHz (NI PXI-5404) is split in two (MiniCircuits splitter ZSC-2-1), with one output being sent to a clock generator board (CGB - AD9959 from Analog Devices) and one to clock a high speed analog to digital convertor (HS ADC - AD9460BSVZ-80 from Analog Devices at 80 M Samples s$^{-1}$). The output from the CGB is controlled via a field programmable gate array card (FPGA - NI PXI-7852R), and sends a sine wave of a desired frequency to some form of modulation device - here an electro-optic modulator (EOM) - to modulate the phase of the input field.\\
Once the modulated signal passes through the cavity, part of it is reflected on to a detector which has two output ports - one ac coupled and one dc coupled. The ac component, containing the modulation signal, is sent to the HS ADC. The dc component, if there is one, is sent to a low speed analog input (AIP $\approx 1$ M Sample s$^{-1}$) of the FPGA. The controller algorithm is discussed in the following section, with the output signal sent through a low speed analog output (AOP), via a high voltage amplifier to the piezo-electric transducer (PZT) controlling the round-trip cavity length.\\
This system has been designed with scalability in mind to cater to experiments with large numbers of locks. Using the same principle as above, up to eight locks can be implemented using two FPGAs, two CGs, one FG, and eight HS ADCs. This is achieved by splitting the 80 MHz FG signal as many times as necessary to send it to clock all HS ADCs and one signal to each CGB (both controlled by one of the FPGAs). Also, the 16 bit signals from pairs of HS ADCs are combined into one 32 bit signal before being sent to an FPGA. This is because the FPGAs cards used here have only two 40 bit digital input/output (DIO) connections, and in this way four locks can be implemented on each FPGA. Each FPGA also has 8 low speed AIPs and 8 low speed AOPs, and another 16 DIO lines all located on a mixed IO connector. These are more than enough for the four dc input and four controller output signals required per FPGA, as well as extra digital lines to control the CGBs.\\
Using this method up to 16 locks can be implemented on one PXI chassis (here a NI PXI-1042Q).
\subsection{Software}
\label{sec:dl_software}
The code was developed using LabVIEW 2010$\textregistered$ (32 bit). An overview of the FPGA code is shown in Fig. \ref{fig:apdl_setup}(b) and consists of two loops: a high speed loop running at 80 M Samples s$^{-1}$; and a low speed loop running at approximately 750 k Samples s$^{-1}$ (limited by the low speed AIP/AOP). In the high speed loop the modulated signal from the HS ADC is demodulated with a cosine function generated using a look-up table at the same frequency as the modulation (i.e. $f_{mod}$). This passes through a Cascaded-Integrator-Comb (CIC) filter \cite{dl7} to produce an error signal that is then sent to the low speed loop.\\
There are two main components to the low speed loop: a lock mode (consisting of a P, I and I$^2$ - PII - controller); and a scan mode (consisting of a saw tooth scan function). Once directed by the user to lock, the code will move from scan mode to lock mode depending on the user defined scan and lock thresholds set for the dc signal. When the user initiates locking, the code will move from scan mode to lock mode when the dc input dips below the scan threshold. The code will then remain in lock mode unless either the user disengages locking or the dc input rises above the lock threshold (set higher than the scan threshold).  In both cases the system will revert to scan mode. In the latter case, however, the system will move back to lock mode once the dc input again drops below the lock threshold. If the lock has no dc component (i.e. a phase-style lock) the absolute value of the error signal can be used to set the thresholds instead.\\
The low speed loop includes a white noise generator, and this is used to measure the system transfer function as described in Sec. \ref{sec:dl_control_theory}. \\
A master code runs on the odd numbered FPGAs and a slave code runs on the even numbered FPGAs. The codes differ in that the master code contains an internal PXI trigger for the FG (only on the first FPGA) and also runs the CGB programming, ensuring that the $f_{mod}$ used for demodulation in the high speed loop is the same frequency as that sent to drive the modulator, and with a fixed phase relation.\\
The real time (RT) code acts as the interface between the FPGAs and the user, as well as to allow communication between FPGAs. This allows for sequential locking logic to be implemented across all FPGAs, with the user specifying the order of the locks, and which locks are dependent on which others. If a lock drops then the controller signal for any locks that depend on the dropped lock will freeze until such time as the first lock is restored. The RT code also includes an inbuilt scope which can display information about any of the locks such as error signal, dc signal and scan signal.
\section{Locking Analysis}
\label{sec:dl_locking_analysis}
In this section we present an experimental analysis of our locking system acting on a bow-tie OPO cavity containing a periodically poled potassium titanyl phosphate (PPKTP) crystal and co-resonant with light at 1064 and 532 nm.
\subsection{Lock Optimization}
\label{sec:lock_optimisation}
\begin{figure}
\begin{center}
\includegraphics[width=\columnwidth]
{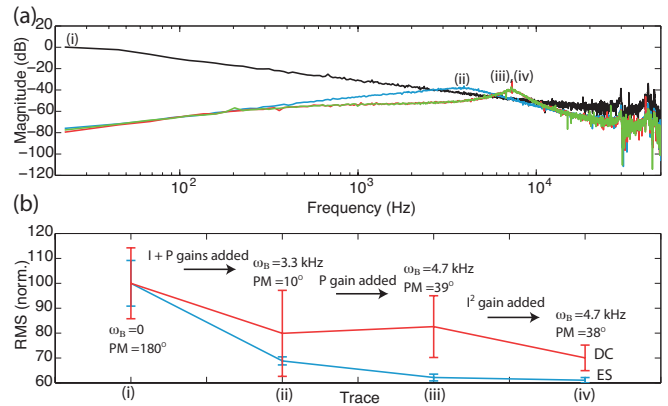} 
\caption{(a) Magnitude plot of system transfer function for an OPO cavity with controller gain increasing from (i) to (iv). (b) Normalised RMS values for these traces for both the error (ES) and dc (DC) signals. Error bars shows standard deviation across 10 runs used for each point. Also shown is the sequence in which the different gains are added, the bandwidth of the system ($\omega_B$) and phase margin (PM) measured by the code.}
\label{fig:lock_optimisation}
\end{center}
\end{figure}
In our analysis two tools were used to investigate the individual locks: the system transfer function and the RMS of the error and dc signals, both measured internally using the code. For the transfer functions, the three key indicators were the amount of noise suppression at low frequencies, the range of frequencies for which noise was suppressed (i.e. bandwidth) and the stability of the lock as measured by the phase margin, as discussed in Sec. \ref{sec:dl_control_theory}.\\
Fig. \ref{fig:dl_controltheory}(b) shows a comparison between the plant transfer function measured using a network analyzer (model MS4630B from Anritsu) as well as using the internal white noise generator. As can be seen, the two match with excellent agreement from approximately 500 Hz onwards, with the frequency range of the white noise generator extending to 100s of kHz. At low frequencies there is extra noise present on the white noise transfer function. Better resolution for these lower frequencies could be achieved by using greater averaging.\\
Fig. \ref{fig:lock_optimisation} illustrates an optimization procedure developed for a generic lock applied to the OPO. Firstly I, and then P, gains are increased iteratively, followed by I$^2$ to increase noise suppression at low frequencies. Fig. \ref{fig:lock_optimisation}(a) shows the system transfer functions for this process and the reduction in noise that occurs at low frequencies. Fig. \ref{fig:lock_optimisation}(b) shows the average RMS for these traces taken over 10 runs, which shows a reduction in noise as more gain is added to the system. Also shown on this figure is the bandwidth and phase margin measured by the system. As can be seen, though the noise suppression seen on the transfer function traces does not increase significantly after the initial I gain is added, the bandwidth does increase while the RMS decreases with increasing gain. In all cases the phase margin was kept above $\pi/6$.\\
Though the amount and ratios of the gains required to optimize each individual lock will differ due to different plant resonances etc., the combination of the P, I, and I$^2$ controllers allow for flexibility when optimizing.
\subsection{Long Term Stability and Comparison with Analog PI Controller}
\label{sec:long_term_stability}
\begin{figure}
\begin{center}
\includegraphics[width=\columnwidth]
{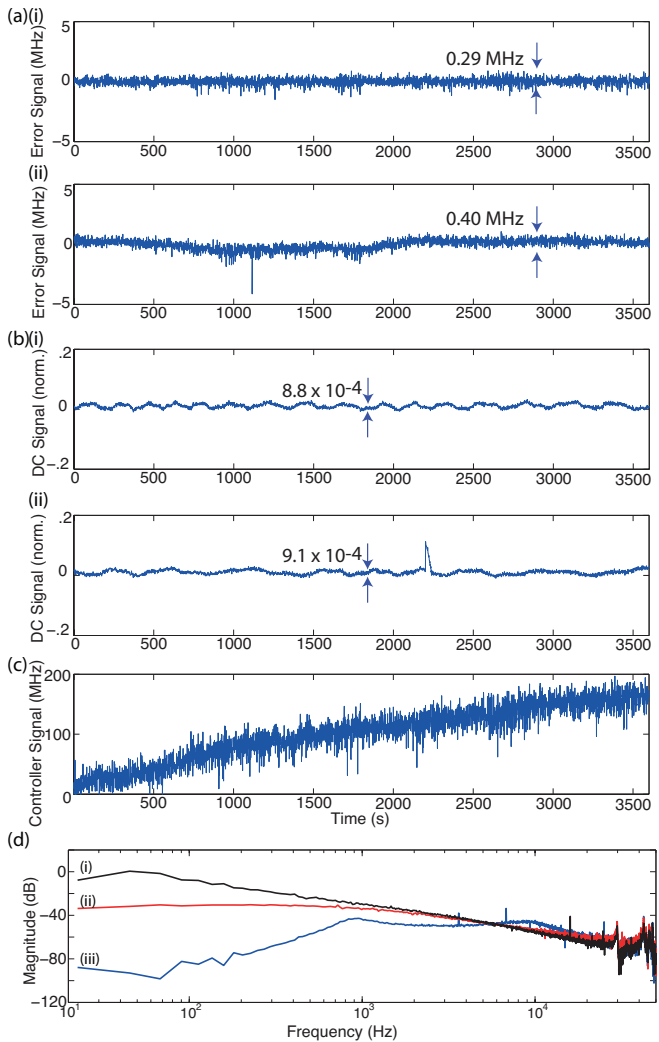} 
\caption{(a) Error signal, measured in MHz, and (b) dc signal, normalized to total peak height, for (i) a digital PII and (ii) analog PI controller, measured over a one hour period. (c) The digital PII controller signal measured during the digital run showing the frequency drift of the laser which was compensated for. (d) Magnitude plots of the system transfer function for (i) no locking, (ii) for locking with the analog PI controller, and (iii) for locking with the digital PII controller.}
\label{fig:dl_stability}
\end{center}
\end{figure}
The long term stability of the OPO cavity was investigated over a period of one hour. To accomplish this, both the error and dc signals for the OPO were measured at 1 second intervals using the code. The OPO was locked firstly with the internally programmed PII controlled and then with an external analog PI controller developed in house. These analog PIs have been used previously on many quantum optics experiments, including the one described in Ref. \cite{bowen}. The results of this comparison are shown in Fig.s \ref{fig:dl_stability}(a) and (b) for the error and dc signals respectively, with the former converted to frequenc and the latter normalized to the maximum peak height.\\
As can be seen, both the analog and digital controllers are stable over the period of one hour, though at one point, approximately 2200 s in to the run, it can be seen that the analog system jumped slightly, but not enough to lose lock. This stability was despite the fact that during an hour the laser resonant frequency was found to move by approximately 175 MHz, measured by monitoring the PII controller output during the digital stability run. This is shown in Fig. \ref{fig:dl_stability}(c).\\
The amount of noise suppression over this time was again measured using transfer functions and RMS values, with the former shown in Fig. \ref{fig:dl_stability}(d). From the RMS values included in Fig.s \ref{fig:dl_stability}(a) it can be seen that the RMS for the digital error signal trace was approximately 0.3 MHz as opposed to the 0.4 MHz for the analog controller. For the dc RMS, even removing the perturbed region around 2200 s from the analog trace, the RMS is approximately the same ($S_i$ being normalized to the total size of the peak and $S_{des}$ is taken to be the peak minima). Also, from Fig. \ref{fig:dl_stability}(d) it can be seen that the digital controller has 50 dB greater noise suppression at low frequencies.\\
These results will vary depending on the different analog and digital controllers developed and the gains used. For instance, the extra noise suppression at lower frequencies for the digital controller can mostly be attributed to the I$^2$ component, which was not included in its analog counterpart. However, it is much easier to alter a digital, rather than an analog, controller and this comparison verifies the efficacy and long term stability of this general digital controller.
\section{Quantum Measurements}
\label{sec:dl_quantum_measurements}
\begin{figure*}
\begin{center}
\centerline{\includegraphics[width=1.8\columnwidth]{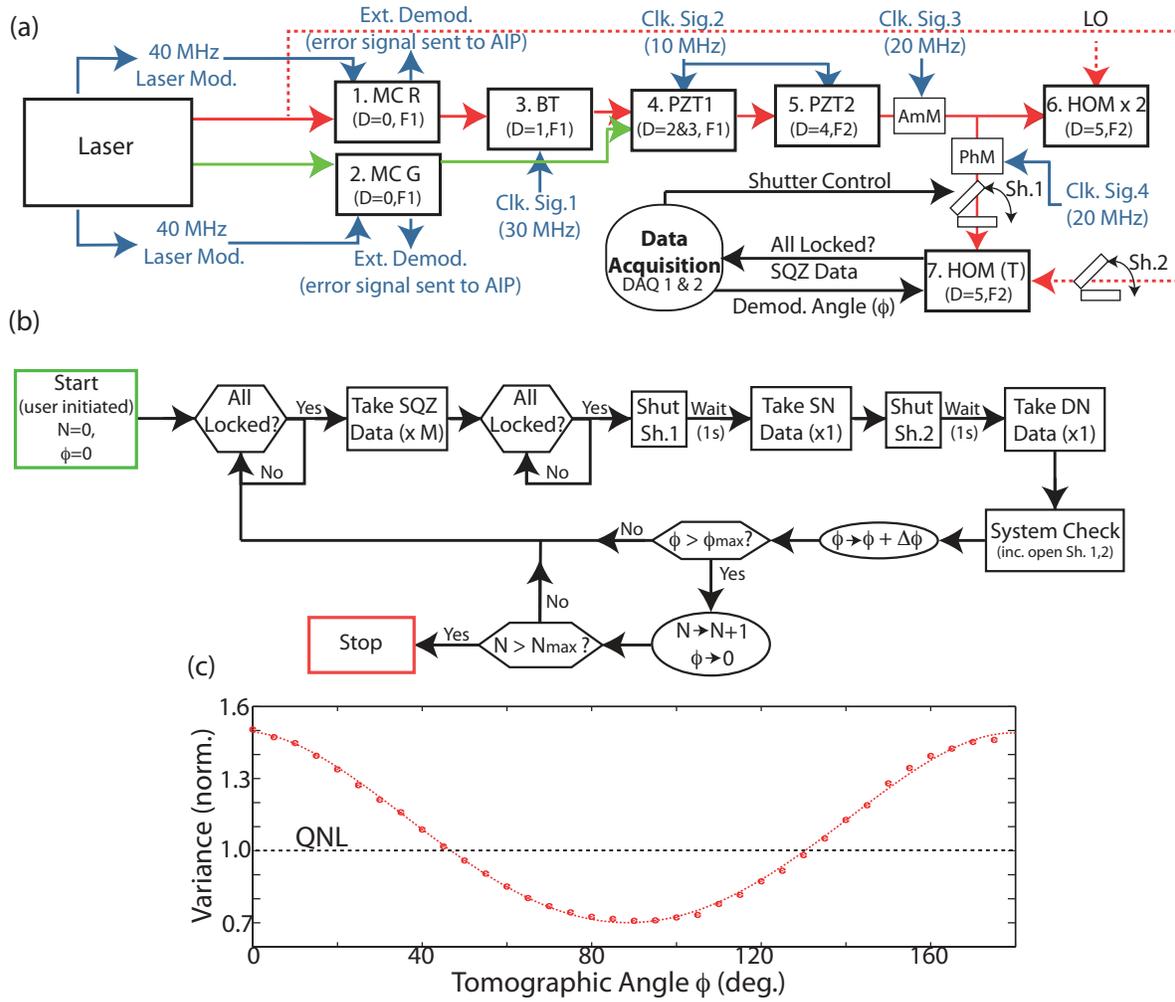}}
\caption{(a) Experimental set-up. MC - modecleaner (R - red, G - green); BT - bow-tie cavitiy; PZT - piezo-transducer phase lock; HOM - homodyne lock (T - tomographic, x 2 - double homodyne); AmM - amplitude modulator; PhM - phase modulator; Sh - mechanical shutter; SQZ Data - squeezing data; D - lock dependencies ($0 = $ no dependencies); F - FPGA lock is controlled from; Ext. Demod. - external demodulation carried out for these locks to produce the error signal; AIP - analog input; DAQ - data acquisition card; Clk. Sig. - clock signal; and LO - local oscillator. (b) Flow diagram of data acquisition procedure used to take squeezing data (SQZ) for various angles of the tomographic lock $\phi$, as well as take dark noise (DN) and shot noise (SN) traces. Values for the number of experimental runs $N$, number of squeezing traces for each tomographic angle $M$, and change in tomographic lock angle between runs $\Delta \phi$  are defined by the user. (c) Plot of variance in noise normalised to the quantum noise limit (QNL), as a function of tomographic lock angle taken using the above procedure with a total of 1.2 million points per angle, and $\Delta \phi = \pi/36$. Circles: data points; dashed line: theoretical fit.}
\label{fig:dl_quantum}
\end{center}
\end{figure*}
So far we have presented results which are not specific to the field of quantum optics. In this section, the OPO is inserted into a larger quantum optics experiment designed to characterize optical Schr\"odinger cat states (see, for instance, Ref. \cite{sc1}). This experiment, described in Ref. \cite{helen} and shown in Fig. \ref{fig:dl_quantum}(a), consists of 2 mode-cleaner cavities (one at 532 nm and one at 1064 nm), the OPO - which is used to produced squeezing on the 1064 nm beam, two phase locks, a double homodyne, and a tomographic homodyne lock to allow locking to any quadrature value. This is all controlled using the hardware and software described in Sec. \ref{sec:dl_code} and integrated into one user-controlled RT code.\\
As in many other quantum optics experiments, it is desirable to only take data when all components of the system are locked. Also, in this case, vast quantities of data (100s of GBs) are required to be able to produce the necessary tomographic reconstructions of the optical Schr\"odinger cats that are characterized, and this is taken over many hours. It is also necessary to compare the squeezing data traces with dark noise (DN - signal and local oscillator blocked) and shot noise (SN - only signal blocked) traces at various times during data collection for normalization purposes.\\
To accommodate these requirements, a second RT code was developed to take data using two data acquisition cards (NI PXI-5124). The data acquisition code communicates with the locking code to determine whether all components are locked, as well as to change locking parameters as desired. For instance, the phase of the single homodyne lock must be changed to allow for the tomographic reconstruction, and blocking of the various fields present using motorized control via the locking FPGA is necessary for collecting DN and SN traces. The data is saved to the hard drive of the PXI controller and later extracted for examination.\\
A flow diagram of the data collection protocol is shown in Fig. \ref{fig:dl_quantum}(b). Here the user determines the number of data traces to take for each tomographic lock angle $M$, as well as the change in tomographic lock angle between measurements $\Delta \phi$, the maximum tomographic lock angle $\phi_{max}$, and the number of experimental runs $N$  for $\phi = 0 \rightarrow \phi_{max}$. It should be noted that the waiting time programmed in to the code is to allow time for the shutters to close or open, and the system check includes such functions as resetting error signal offsets so that all locks remain stable during the hours of data collection. If a lock does fail during the measurement of one angle, the protocol will reset and try again for that angle until it is successful. Fig. \ref{fig:dl_quantum}(c) shows an example of squeezing data collected from one such run with $\Delta \phi = \pi/36$. Approximately 1.5 dB of noise suppression below the quantum noise limit (QNL) can be seen for a tomographic angle of $\phi=90^{\circ}$, accompanied by an increase of 1.8 dB in the opposite quadrature.\\
This implementation of a digital control algorithm on a uniquely quantum system highlights the benefits digital control can bring to quantum optics experiments: the integration of information allowed for data to only be taken when all system components were locked to ensure only useful data was collected; the automation of the locks meant that the experiment could be left unattended for many hours to collect the required amount of data as quickly as possible; and the added flexibility allowed for automated acquisition of SN and DN, as well as the ability to alter the homodyne angle to perform the tomographic measurements.\\
\section{Conclusions}
\label{sec:dl_conclusions}
In conclusion, we have presented a digital locking system that allows for automatic and sequential locking and is easily scalable. This code was programmed using LabVIEW$\textregistered$ software and is free to download. We have shown how the inbuilt locking analysis tools can be used to help optimize individual locks, as well as demonstrated the efficacy and long term stability of the digital controller. Finally, we used an example of a experiment used to characterize optical Schr\"odinger cat states to illustrate the advantages of digital control for quantum optics experiments.
\section{Acknowledgements}
This research was conducted by the \textit{Australian Research Council Centre of Excellence for Quantum Computation and Communication Technology} (project number CE110001027).

\end{document}